\documentclass[acmtog]{acmart}

\AtBeginDocument{%
  \providecommand\BibTeX{{%
    \normalfont B\kern-0.5em{\scshape i\kern-0.25em b}\kern-0.8em\TeX}}}

\setcopyright{acmlicensed}
\copyrightyear{2024}
\acmYear{2024}
\acmDOI{XXXXXXX.XXXXXXX}

\usepackage{lipsum}
\usepackage{xcolor}
\usepackage{cleveref}
\usepackage{lineno}
\usepackage{subfigure}
\usepackage{graphicx}
\usepackage{subcaption} 
\usepackage{booktabs}   
\usepackage[aboveskip=5pt, belowskip=0pt]{caption} 

\begin{document}


\title{2-Factor Retrieval for Improved Human-AI Decision Making in Radiology}

\author{Jim Solomon}
\authornote{These authors contributed equally to this research.}
\email{jimsolomon@ucla.edu}


\author{Laleh Jalilian}
\authornotemark[1]
\email{ljalilian@mednet.ucla.edu}

\author{Alexander Vilesov}
\authornotemark[1]
\email{vilesov@ucla.edu}

\author{Meryl Mathew}
\email{merylmathew@ucla.edu}

\author{Tristan Grogan}
  \email{tgrogan@mednet.ucla.edu}

\author{Arash Bedayat}
\email{abedayat@mednet.ucla.edu}

\author{Achuta Kadambi}
\affiliation{%
  \institution{University of California, Los Angeles}
  \city{Los Angeles}
  \country{USA}}
\email{achuta@ee.ucla.edu}

\begin{abstract}

Human-machine teaming in medical AI requires us to understand to what degree a trained clinician should weigh AI predictions. While previous work has shown the potential of AI assistance at improving clinical predictions, existing clinical decision support systems either provide no explainability of their predictions or use techniques like saliency and Shapley values, which do not allow for physician-based verification. To address this gap, this study compares previously used explainable AI techniques with a newly proposed technique termed `2-factor retrieval (2FR),' which is a combination of interface design and search retrieval that returns similarly labeled data \emph{without processing} this data. This results in a 2-factor security blanket where: (a) correct images need to be retrieved by the AI; and (b) humans should associate the retrieved images with the current pathology under test. We find that when tested on chest X-ray diagnoses, 2FR leads to increases in clinician accuracy, with particular improvements when clinicians are radiologists and have low confidence in their decision. Our results highlight the importance of understanding how different modes of human-AI decision making may impact clinician accuracy in clinical decision support systems.

\end{abstract}



\received{20 February 2007}
\received[revised]{12 March 2009}
\received[accepted]{5 June 2009}

\maketitle


\section{Introduction}

In medicine, barriers to the acceptance of artificial intelligence (AI) tools in clinical workflows occur due to the "black box" nature of AI, which make it difficult for clinicians to understand and trust the predictions of a model~\cite{feldman2019artificial,miller2019explanation, ribeiro2016should}. A popular method to elucidate a model's decision is to include explanations in the form of textual or visual elements to help clarify how the components of a given image, such as specific areas in an image, influence a model’s prediction. For instance, a doctor may make more informed clinical decisions when a model offers clear and understandable explanations of its predictions, and a lack of understanding of the prediction's rationale could hinder clinicians from identifying and addressing errors, particularly in scenarios when model prediction and clinician intuition are discordant ~\cite{naiseh2023different}. 

\begin{figure}[t]
  \includegraphics[width=0.45\textwidth]{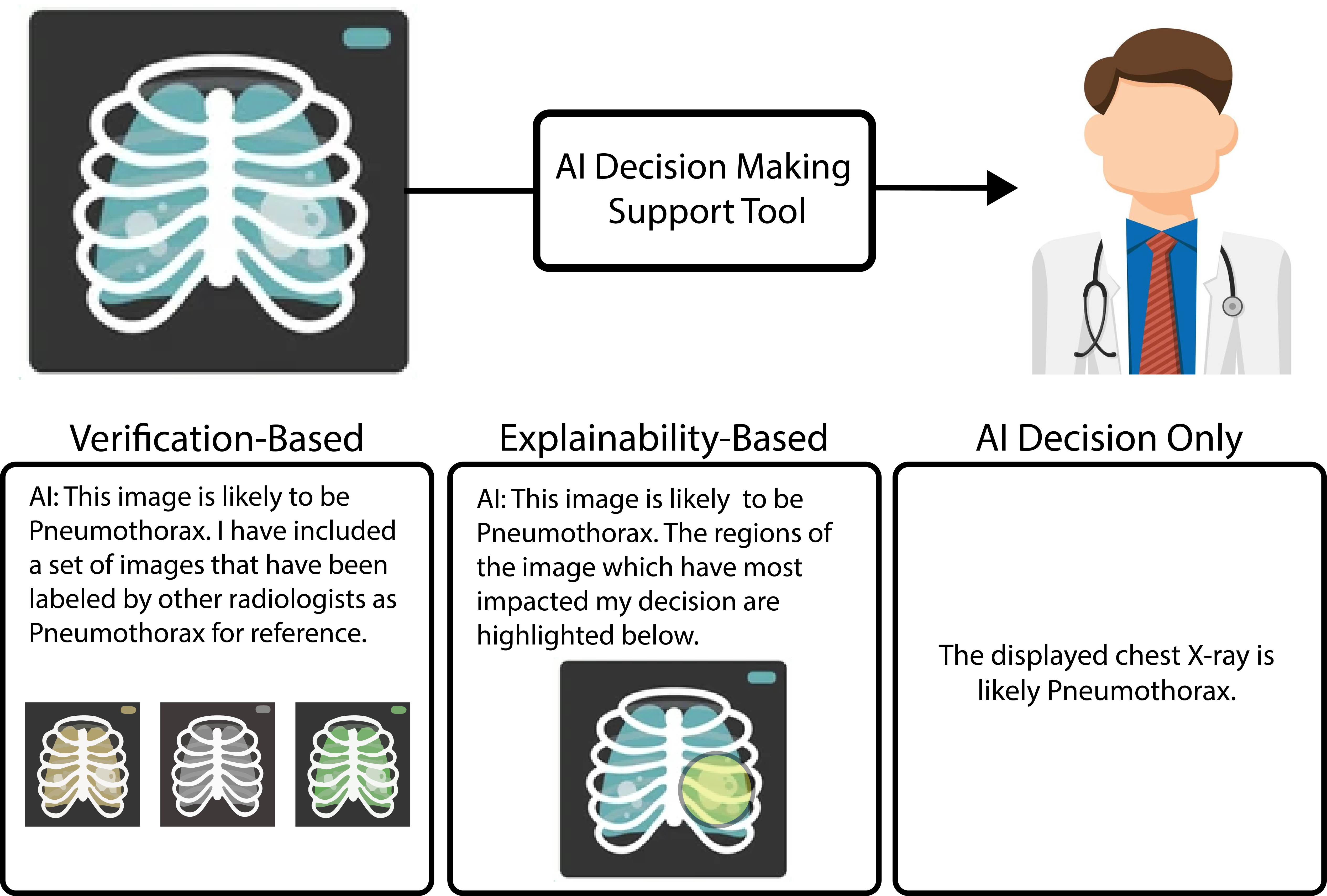}
  \caption{Different Modes of AI-Human Decision Making, including Verification-Based, Explainability-Based, and AI-Decision Only.}
  \label{fig:teaser}
\end{figure}

However, prior work has shown that model explainability can lead to users falsely trusting an AI model's decision due to convincing explanations of incorrect decisions~\cite{jacobs2021machine}. While the field has focused on developing models that can explain the reasoning behind a particular diagnosis or treatment recommendation, for example showing which factors or variables are most important in a model’s prediction, a knowledge gap exists in understanding whether other aspects of AI-human decision may offer distinct advantages over another and how users account for these modes of decision-making in their acceptance of AI predictions. Specifically, we study the impact of verification-based AI-human decision making. Verification-based AI-human decision making encourages human's to attempt to verify an AI prediction before accepting the decision. To facilitate this, we introduce a simple technique that can be paired with any AI prediction tool that encourage a human evaluator's recall and verification abilities. The newly proposed technique termed `2-factor retrieval (2FR)' is a combination of interface design and search retrieval that returns similarly labeled data. This results in a 2-factor reasoning step where: (a) correct images are retrieved by the AI, and (b) humans should associate the retrieved images with the current pathology under test. Given an AI predicted diagnosis for an image, we present the evaluator with canonical image examples of the given AI diagnosis. This method allows the clinician to recall the salient features of the diagnosis and furthermore compare the canonical images with the current image. We evaluate the efficacy of our proposed method on a diverse set of clinicians with varying expertise and years of experience and compare against other modes of AI-human decision making. We present the following contributions of this work:

\begin{enumerate}
    \item We evaluate how various modes of AI-human decision making impacts clinician confidence in their diagnosis through a clinical study on AI-assisted chest x-ray diagnosis.
    \item We perform a comprehensive analysis across different variables providing insights into how various modes of AI-human decision making affects clinician accuracy as a function of the difficulty of the problem, whether the AI was correct, and other clinician variables such as expertise and years of experience. 
    \item We introduce a new technique for verification-based AI, 2FR, which allows users to compare the AI prediction with similarly labeled images. We find that '2FR' outperforms other modes of AI-human decision making that were included in our analysis.
\end{enumerate}

\section{Related Works}

\subsection{Explainable AI}
\label{ssec:XAI}

With the increasing interest to implement machine learning into real-world clinical settings, the role of explainable and interpretable machine learning has been one way in understanding if AI can facilitate more informed and accurate decision making~\cite{ehsan2021operationalizing, bilgic2005explaining}. Early efforts in explainable AI (XAI) focused on feature-based explanations~\cite{rajpurkar2017chexnet,selvaraju2017grad}. Recent considerations of interpretability comprise a wider range of techniques, including uncertainty and confidence metrics~\cite{schaekermann2020ambiguity}, nearest-neighbors~\cite{henry2022human}, and counterfactuals~\cite{zytek2021sibyl}. 

However, there have been mixed results on whether explanations actually help clinicians who are making AI-supported decisions~\cite{ghassemi2021false}. The literature demonstrates that individuals tend to be swayed by AI, frequently accepting its decisions without proper verification, a phenomenon termed overreliance~\cite{buccinca2021trust}. Among various error types observed in human-AI decision-making, overreliance emerges as the most common issue identified in empirical studies. This tendency involves individuals ceding their decision-making responsibility and accountability to AI systems, which will be problematic in critical domains like healthcare. Such overreliance not only risks amplifying machine biases but does so under the pretense of human intervention and control. Additionally, models can also generate seemingly sensible explanations for incorrect predictions~\cite{bussone2015role}. 

\subsection{AI-assisted Decision-making}

With the advancements in AI models, we have seen an explosion of research into their applications as well as early adoption of the technology into industry. In healthcare, AI is viewed as a technology meant to augment clinicians in the quality of their decision making and not replace them. Such a collaboration between AI and humans requires understanding the explainability needs of end-users, in order to best develop appropriate reliance between the clinician and the AI. As an example, does the optimal interface require just the AI's answer, or do clinicians require some level of interpretibility to the model's predictions?  A large area of scholarship focuses on advancing methods of explaining the decision-making process of the AI as summarized in \cref{ssec:XAI}; however, it is equally important to understand to what extent do explainability methods help in improving performance~\cite{horne2019rating, lai2019human, lai2020chicago} or in some cases even hurt performance~\cite{buccinca2020proxy, buccinca2021trust}. This research avenue has confirmed the potential of error in human reasoning such as confirmation bias~\cite{wang2019designing, kim2022hive, buccinca2021trust}, further worsened by anchor bias~\cite{wang2019designing, ghai2021explainable, nourani2021anchoring}, as well as increased confidence in decision despite no correlation with accuracy~\cite{alam2021examining, sivaraman2023ignore, kim2022hive}. Confirmation bias is becoming increasingly worrisome due to the ability of large language models (LLMs) to write convincing text even when the outputs are factually incorrect~\cite{si2023large}. \citet{fok2023search} found in a survey that the majority of AI applications do not yield complementary performance when explanations are included unless the explanation helps verify the accuracy of the answer. \citet{vasconcelos2023explanations} found that extra care is required in forming the explanation to reduce the likelihood of overreliance in AI systems which is a common issue that has been identified in using explainable AI. 

\subsection{AI Decision Support in Medicine}

Improvements in AI diagnostic abilities have led researchers to explore methods that incorporate a model's prediction into clinicians' workflow. The majority of research has focused on creating AI decision support frameworks where AI augments human decision making~\cite{khosravi2024artificial}. Other works have focused on the proper presentation of AI decisions within clinical workflow such as data visualization, risk presentation, communication of system properties and other design considerations~\cite{yang2019unremarkable, timmermans2004different, tait2010effect, sultanum2018more, cai2019hello}. Several studies within this area have pointed out that adoption of AI decision support systems is low~\cite{devaraj2014barriers, elwyn2013many}, especially in prognostic focused applications, thus reducing the ability to analyze their effectiveness~\cite{wyatt1995commentary}. However, ~\citet{scheetz2021survey} performed a survey on clinicians preferences for such systems and found positive attitudes towards how AI could affect their workflow in increasing accuracy and reducing time. Despite low adoption, works have found utility in AI decision support systems increasing diagnostic accuracy and reducing time spent on repetitive tasks. In dermatology, simple merging of human and AI decisions have been shown to increase accuracy~\cite{hekler2019superior, tschandl2020human}, ~\cite{barata2023reinforcement} used reinforcement learning to adjust the risk-reward of AI model decisions in clinician support systems to better represent human preferences, and ~\cite{groh2024deep} studied how fairness in accuracy of AI systems across skin tones impacts clinicians' overall diagnostic performance. In radiology, ~\citet{xie2020chexplain} conducted an iterative design of a support system based on clinician feedback, ~\cite{yu2024heterogeneity} found improvements in clinician accuracy while ~\cite{gaube2023non} found that non-radiologist clinicians benefited most from AI input, and ~\cite{gaube2021ai} explored how incorrect AI decisions and explanations affected clinician decision making, showing evidence of over-reliance. Similar work has been conducted in other fields with most concentrating on ophthalmology~\cite{li2023artificial, guo2024improve}, cardiology~\cite{massalha2018decision, olawade2024advancements}, and neurology~\cite{shahtalebi2021deep, gombolay2024effects}.

\section{Study Methods}

\subsection{Research Aim}

The purpose of this study was to assess the utility of different methods of AI-human decision making. We assess these systems in joint AI-human interpretations of clinical Chest X-Rays. We compare different modes of presenting AI predictions and their influence on clinician accuracy and confidence. 

\subsection{Participants}

In total, N = 69 participants finished the online experiment and were included in the data analysis. The sample consisted of physicians with different levels of task expertise and different years of training. Physicians trained in internal medicine, anesthesiology, surgery, or emergency medicine often review chest X-rays but, compared to Radiology, have relatively little formal training in viewing medical images and were consequently classified as "non-task experts". Radiologists with specialized training in reviewing medical images were classified as "task experts." Participants were recruited via email. Study invitations were sent to staff and residents at hospitals in the US and to residency program coordinators with the request to distribute the link. Table \ref{tab:fig_one_table} displays the participant demographics.

\begin{table}[t]
    \centering
    \begin{tabular}{p{0.25\linewidth} | p{0.2\linewidth} p{0.2\linewidth} p{0.2\linewidth}} 
        \toprule
        Characteristics & Radiologists, \newline (n = 25) & Non \newline Radiologists (n = 44) & Total \newline Respondents (n = 69) \\ [0.5ex] 
        \midrule \midrule
         \textbf{Sex} &  &  & \\ 
         Male & 18  & 19 & 37\\
         Female & 7 & 23 & 30\\
         Other & 0 & 2 & 2\\
        \midrule
         \textbf{Race} &  &  & \\
         White & 14 & 20 & 34 \\
         Asian & 6 & 14 & 20 \\ 
         AIAN & 0 & 1 & 1 \\
         Black & 2 & 5 & 7 \\
         NHPI & 0 & 2 & 2 \\
         Other & 3 & 2 & 5 \\
         \midrule
         \textbf{Years in \newline Practice} &  &  & \\
         0 - 10 Years & 19 & 26 & 45 \\
         11+ Years & 6 & 18 & 24 \\ [1ex]
        \bottomrule
    \end{tabular}
    \caption{Participant Characteristics and Demographics. Non-radiologists include Anesthesiology, Internal Medicine, Emergency Medicine, Surgery. NHPI means Native Hawaiian/Pacific Islander and AIAN means American Indian/Alaskan Native. }
    \label{tab:fig_one_table}
\end{table}

\subsection{IRB Approval}

The UCLA Institutional Review Board (IRB) approved the study, and informed consent was obtained from all participants. This research complies with all relevant ethical regulations. The UCLA Institutional Review Board approved this study as IRB Exempt. At the beginning of the experiment, all participants were presented with the following informed consent statement: "CXR Diagnosis is a UCLA research project. All submissions are collected anonymously for research purposes. You can leave this website anytime."

\begin{figure}[t]
    \centering
    \noindent\includegraphics[width=0.45\textwidth]{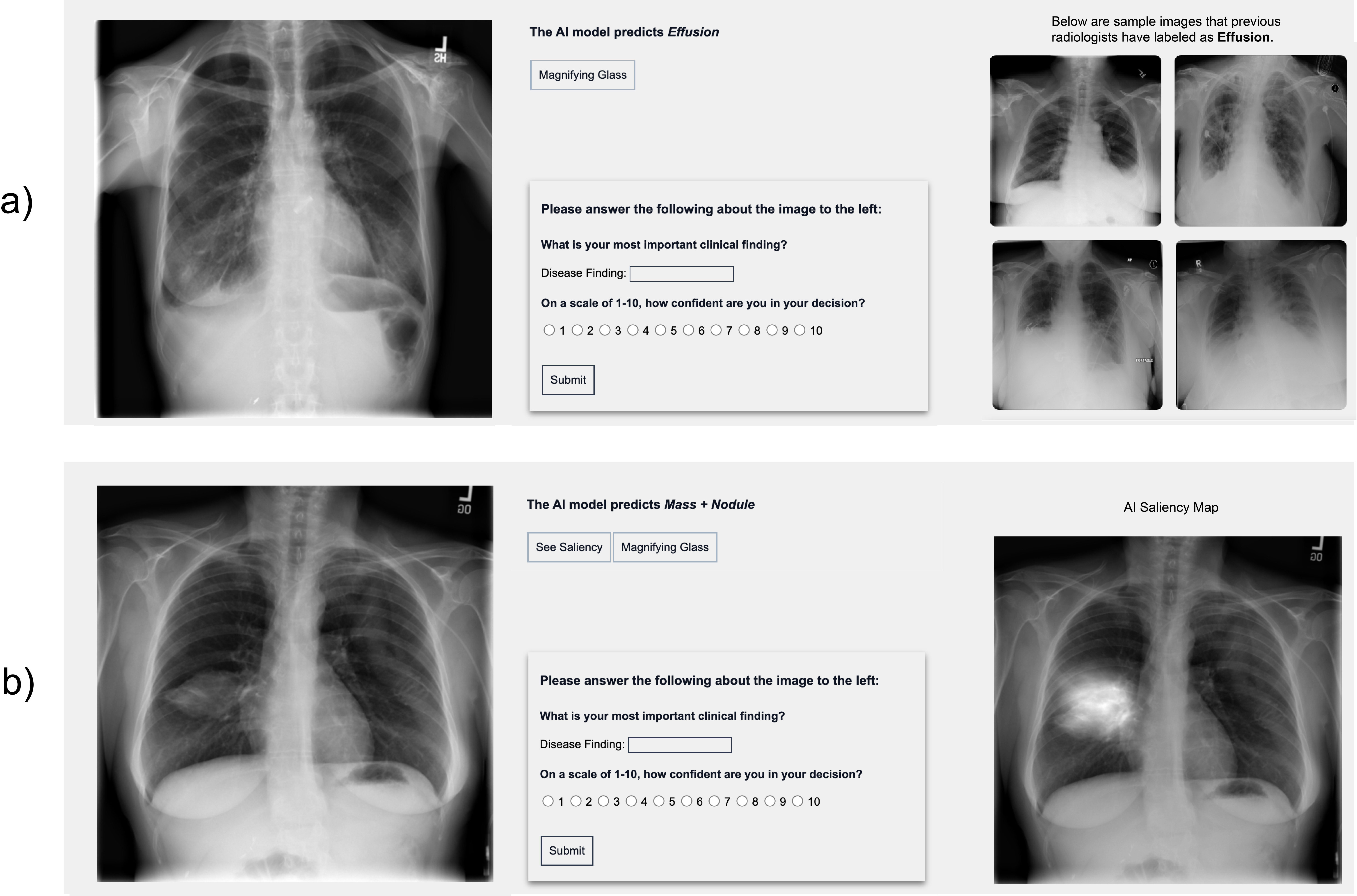} 
    \caption{Example interface shown to radiologists. Panel A demonstrates 2FR, where four images with physician-confirmed pathology are retrieved and used as canonical examples of the AI predicted pathology. Panel B shows a Saliency map, where a section in the image is identified as displaying the AI predicted pathology.}
    \label{fig:sample-survey}
\end{figure}

\subsection{Experimental Design}

The experiment was conducted online via a publicly accessible website which we developed. An example interface shown to radiologists can be seen in Fig. \ref{fig:sample-survey}. Participants were given basic information about the purpose of the study and an estimated study duration of 10 to 15 min. They were informed that participation was completely voluntary and anonymous, that they could quit the study at any time without negative consequences, and about the option of being included in a raffle as compensation for their participation. Only individuals who gave written informed consent to take part in the study (by clicking a checkbox) and confirmed that they were currently practicing radiology, internal medicine, anesthesiology, surgery, or emergency medicine (residency included) in the USA or Canada could move on to the experiment. Participants completed a short survey, including questions about demographics, professional identification, and years of experience.

The remaining 12 questions were designed to determine whether different modes of explainability impact clinician confidence in AI-assisted predictions. 24 images were taken from the NIH Chest X-ray dataset \cite{wang2017chestx}. The paper introducing the Chest X-ray dataset included a deep convolutional neural network (DCNN) that predicts a pathology in the chest X-ray images and provides saliency maps that explains which regions of the X-ray image contribute most to the AI's decision. We used the saliency maps from ~\cite{wang2017chestx} as a benchmark for explainability-based methods. To understand human-AI decision making behavior when the AI is correct and incorrect, we manually selected 2/3 of presented instances to be when the DCNN was accurate and the rest are when the DCNN was inaccurate. We utilized the model's label predictions and the saliency maps for our experiments. We split the images randomly into two sets for survey versions A and B, and each chest X-ray presented one of four conditions: Mass/Nodule, Cardiomegaly, Pneumothorax, or Effusion. Both sets contained three images of each condition, and half of the chest X-rays had a diagnosis difficulty rating of "Easy" while the other half was labeled "Hard."

Participants were assigned a random survey version and random ordering of the images in the corresponding set. Each question presented an image and participants were given 14 options to choose from in diagnosing it. The image was accompanied by either an AI diagnosis, an AI diagnosis with the option to view the highlighted salient regions of the image to the prediction (Saliency), an AI diagnosis along with four more images recognized by other physicians to represent that diagnosis (2FR), or no AI assistance at all. Each survey contained three questions of each AI modality in total, though they were randomly assigned to the 12 images. The overall AI prediction accuracy rate in both survey versions was 66.67 percent, which was deliberately low to better ascertain the disparities in diagnostic accuracy and confidence between AI-assisted predictions and ones made without AI input.

\subsection{Statistical Analyses}

To determine whether confidence levels varied by modality, we conducted linear mixed-effects models with fixed effects for AI modality, question difficulty, participant specialty, years of practice, and participant age. A random intercept was included to account for within-participant variability. Similar models were constructed for accuracy. Least squares means (LSMeans) were used to compare confidence levels across AI modalities, with pairwise differences and 95 percent confidence intervals estimated. Analyses were conducted using SAS V9.4 (Cary, NC) and p-values <0.05 were considered statistically significant.

\subsection{Procedure}

In the survey, participants learned that their task was to review and diagnose 12 patient cases as accurately as possible, for which they received chest X-rays and the diagnostic advice that could be used for their final decisions. The chest X-rays were shown as a static image on the survey site. For each case, the participating physicians were asked to pick a diagnosis and judge how confident they were with their diagnosis. 

\subsection{Measures}

The present study had two dependent variables: (1) diagnostic accuracy, and (2) confidence in the diagnosis.

Diagnostic accuracy: After being presented with the AI-generated diagnosis, the participating physicians were asked "What is your most important clinical finding?" to provide their own diagnosis from a limited set of options, without being prompted to explicitly agree or disagree with the AI prediction. The accuracy of the physician’s diagnosis was determined by comparing their selected diagnosis with the correct diagnosis associated with each case. Since the AI diagnosis was correct approximately two-thirds of the time, a tertiary variable was also analyzed: the alignment or correlation between physician’s diagnoses and the AI-generated diagnoses. This alignment was used to explore how often physicians followed the AI's advice and therefore their confidence in AI predictions as a whole.

Confidence in the diagnosis: For each case, participants rated the confidence in their final diagnosis with one item ("How confident are you with your primary diagnosis?") on a 10-point Likert scale from 1 (not at all) to 10 (extremely).

\section{Results}

\subsection{Accuracy Across Modes of AI-Human Decision Making}

\begin{figure}[t]
    \centering
    \includegraphics[width=0.45\textwidth]{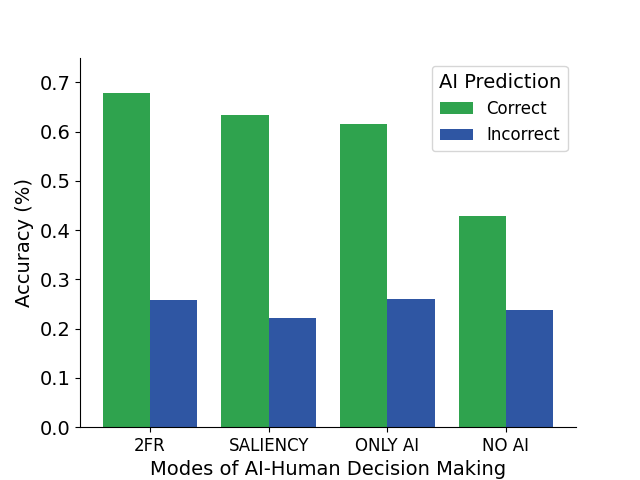}
    \caption{Physician accuracy across AI correctness.}
    \label{fig:figure_acc_ai_correctness}

    \vspace{0.5cm}

    \begin{tabular}{p{0.4\linewidth} p{0.1\linewidth} p{0.1\linewidth} p{0.1\linewidth} p{0.1\linewidth}} 
    \toprule
    & Mean & SE & Lower 95\% CI & Upper 95\% CI \\ [0.5ex] 
    \midrule
     2FR v AI Correct & 0.69 & 0.06 & 0.57 & 0.81 \\
     2FR v AI Incorrect & 0.27 & 0.07 & 0.12 & 0.43 \\ 
     Saliency v AI Correct & 0.65 & 0.06 & 0.52 & 0.77 \\
     Saliency v AI Incorrect & 0.25 & 0.07 & 0.11 & 0.39 \\ 
     AI Correct & 0.64 & 0.06 & 0.51 & 0.76 \\
     AI Incorrect & 0.27 & 0.07 & 0.13 & 0.42 \\          
     No AI v AI Correct & 0.45 & 0.06 & 0.33 & 0.58 \\ 
     No AI v AI Incorrect & 0.24 & 0.07 & 0.10 & 0.39 \\ [1ex] 
    \bottomrule
    \end{tabular}
    \caption{Standard error and confidence intervals of physician accuracy across AI correctness.}
    \label{tab:figure_acc_ai_correctness}
\end{figure}

One of our main experimental goals was to understand how various modes of AI-Human decision-making impacts clinician accuracy. For this, we recruited $N = 69$ physicians to participate in a survey. The result for mean accuracy across modalities are shown in Fig. \ref{fig:figure_acc_ai_correctness}, while Fig. \ref{tab:figure_acc_ai_correctness} shows the same accuracy values, coupled with its standard error and confidence intervals.

\noindent
\paragraph{Overall}

Across all modes where AI assistance is provided, physician accuracy is markedly higher when AI predictions are correct (0.35 (95\% CI 0.28-0.41), p<0.001). The impact of AI being correct or not on accuracy does not significantly vary by modality type (e.g. 2FR, Saliency, AI, no AI). In the 2FR modality, physicians achieve the highest overall accuracy (~70\%), suggesting that providing physicians with AI-predicted diagnoses alongside the 2FR metric enhances AI-Human decision making. A similar trend is observed in the AI Saliency modality, where accuracy remains high (~65\%), indicating the utility of providing visual or contextual cues to support AI predictions. In contrast, when physicians rely solely on AI predictions without supplemental information (AI modality), their accuracy is slightly reduced (~64\%). This finding underscores the limitations of using AI outputs in isolation, which may not fully inform human decision making. In the absence of AI assistance (No AI modality), physician accuracy declines further (~45\%), underscoring the critical role of AI systems in augmenting human performance in decision making tasks. In cases where AI predictions are incorrect, physician accuracy across all modalities is substantially lower, with minimal variation between 2FR, Saliency, Only AI and No AI. This suggests that when the AI is wrong, clinicians rely on their own expertise. From Fig. \ref{fig:figure_acc_ai_correctness} and Fig. \ref{tab:figure_acc_ai_correctness}, we see that AI correctness significantly influences the performance of the AI-Human decision making (p<0.001), suggesting that physicians are overly trusting of AI predictions.

We notice in Fig. \ref{fig:figure_acc_ai_correctness} that AI-Human accuracy is lower when AI is incorrect, even when there is no AI prediction being served to a clinician. This could be associated with task complexity. Cases associated with incorrect AI predictions might inherently be more difficult, skewing results. Without AI assistance, physicians face these difficult cases alone, performing poorly on them due to its inherent difficulty.

\begin{figure}[t]
    \begin{subfigure}{
        \includegraphics[width=0.9\linewidth]{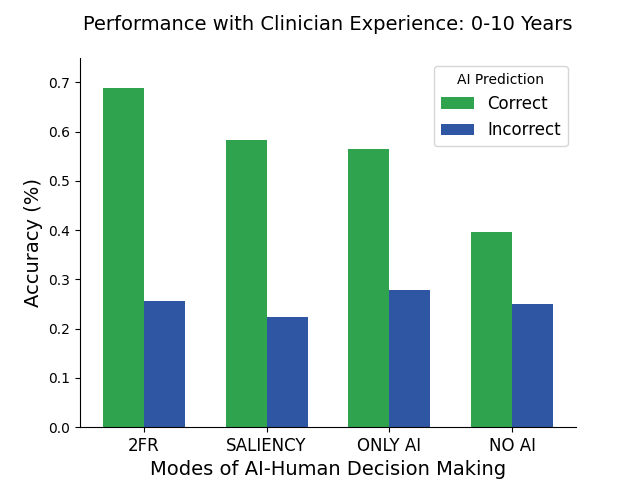}
    }
    \end{subfigure}
    \vfill   
    \begin{subfigure}{
        \includegraphics[width=0.9\linewidth]{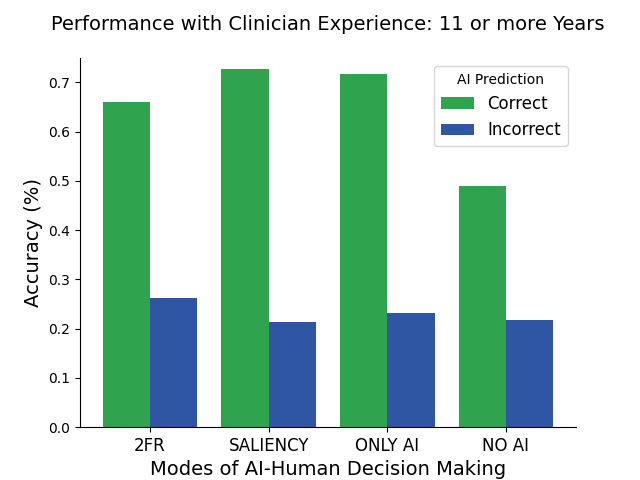}
    }
    \end{subfigure}
    \caption{Accuracy across modes of AI-Human decision making and AI correctness based on clinician experience.}
    \label{fig:fig_acc_expertise}
\end{figure}

\noindent
\paragraph{Experience}

Fig. \ref{fig:fig_acc_expertise} highlights the influence of clinical experience on the effectiveness of AI-Human decision-making. When clinicians have less than 11 years of experience, 2FR achieves the highest accuracy when AI predictions are correct. The accuracy reaches approximately 70\%, while for those with 10 or more years, it slightly declines to around 65\%. This suggests that 2FR is most useful for clinicians with less experience. Across all modes, incorrect AI predictions lead to substantial performance declines. However, the accuracy values are comparable to the No AI, suggesting that clinicians rely more on their expertise when the AI is incorrect.

\noindent
\paragraph{Expertise}

\begin{figure}[t]
    \begin{subfigure}{
        \includegraphics[width=0.9\linewidth]{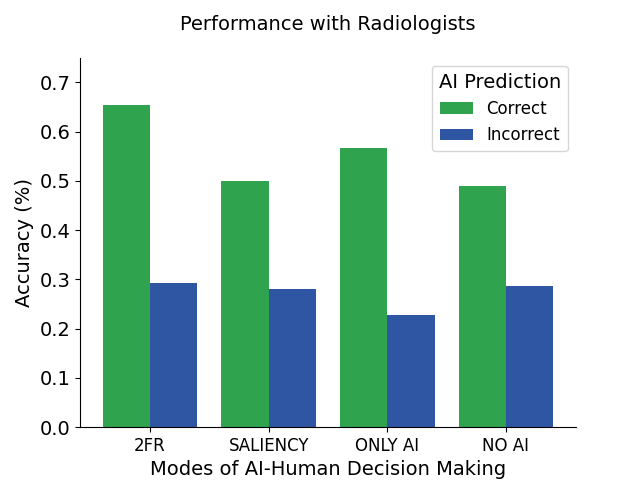}
    }
    \end{subfigure}
    \vfill   
    \begin{subfigure}{
        \includegraphics[width=0.9\linewidth]{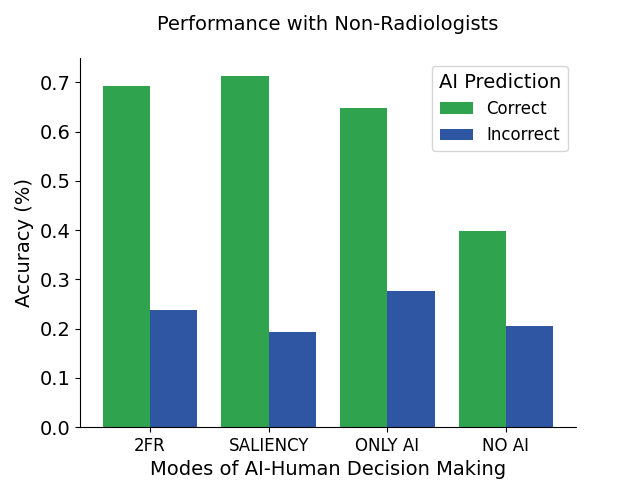}
    }
    \end{subfigure}
    \caption{Accuracy across modalities and AI correctness based on clinician expertise.}
    \label{fig:fig_acc_specialty}
\end{figure}

In Fig. \ref{fig:fig_acc_specialty}, radiologists using the 2FR modality achieve the highest accuracy when AI predictions are correct (~65\%). This demonstrates that incorporating 2FR metrics into AI assistance is highly effective for expert users. Scheetz \cite{scheetz2021survey} showed that radiologists have a high standard for AI correctness and prefer using AI to automate monotonous tasks. This explains the result where 2FR performs best when AI is correct. The questions in which the AI is correct can be interpreted as easy questions. This makes them a more monotonous task to a radiologist. Saliency and Only AI also yield good performance (~50\% and ~60\%, respectively) when AI predictions are correct, though lower than 2FR. In the case of Non-radiologists, the difference between the accuracy of 2FR and Saliency is marginal. This suggests that 2FR significantly aids expert and non-expert clinicians, but AI Saliency harms expert users. We observe a 20 point drop in accuracy from non-expert to expert with the Saliency modality, this suggests 2FR is a more robust modality across clinician expertise. The comparison reveals that radiologists, despite their domain expertise, benefit significantly from AI assistance, particularly when provided with supportive features such as 2FR. However, they are less reliant on AI and more resilient to errors compared to non-radiologists. Non-radiologists show greater dependence on AI outputs and are more vulnerable to incorrect predictions.

\paragraph{Chest X-Ray Difficulty}

\begin{figure}[t]
    \begin{subfigure}{
        \includegraphics[width=0.9\linewidth]{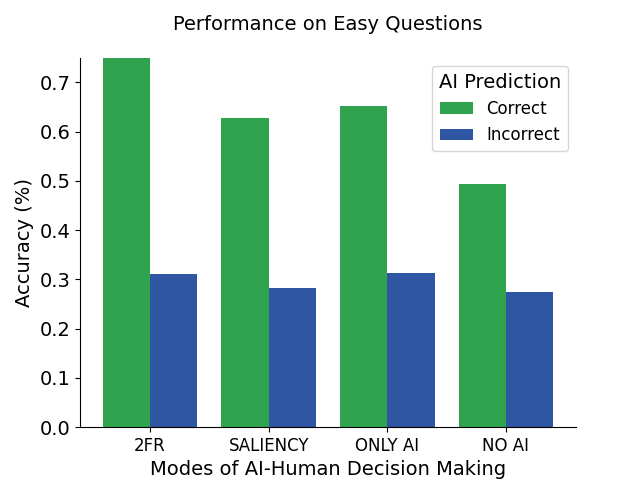}
    }
    \end{subfigure}
    \vfill   
    \begin{subfigure}{
        \includegraphics[width=0.9\linewidth]{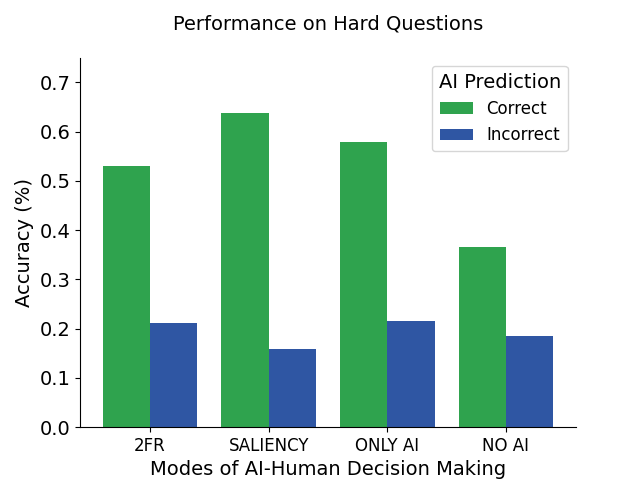}
    }
    \end{subfigure}
    \caption{Accuracy across modalities and AI correctness based on chest x-ray difficulty.}
    \label{fig:fig_acc_correctness}
\end{figure}

Fig. \ref{fig:fig_acc_correctness} reveals distinct trends in performance for easy versus hard chest X-ray cases. For easy questions, all modalities show a significant advantage when AI predictions are correct, with the 2FR yielding the highest accuracy (>70\%). We see higher accuracy on 2FR on easier and correct questions, implying that 2FR assists clinicians in more accurate diagnoses. For hard questions, accuracy decreases across all modalities except Saliency. The accuracy of Saliency remains consistent when AI prediction is correct across Easy and Hard questions.

\paragraph{Reliance On AI}

With p < 0.001, we observe a significant correlation between clinician accuracy and AI correctness across all AI-Human decision-making modalities. When the AI prediction is correct, clinician accuracy is substantially higher across 2FR, Saliency, and Only AI modalities compared to the No AI condition. This demonstrates that clinicians leverage AI effectively when it provides accurate information, enhancing their diagnostic performance.

However, when the AI is incorrect, the difference in accuracy between the AI-assisted modalities and the No AI condition becomes marginal. For example in Fig. \ref{fig:figure_acc_ai_correctness}, clinician accuracy in 2FR, Saliency, and Only AI modalities (~25\%) is similar to the accuracy in the No AI condition (~25\%). This minimal difference suggests that clinicians do not heavily rely on AI predictions when they are incorrect. Instead, they appear to fall back on their own expertise and experience, resulting in comparable performance to the No AI scenario. Furthermore, this trend is observed irrespective of clinician expertise and experience levels, indicating a generalized behavior across the clinical population.

\subsection{Clinician Confidence}

\begin{figure}[t]
    \centering
    \noindent\includegraphics[width=0.45\textwidth]{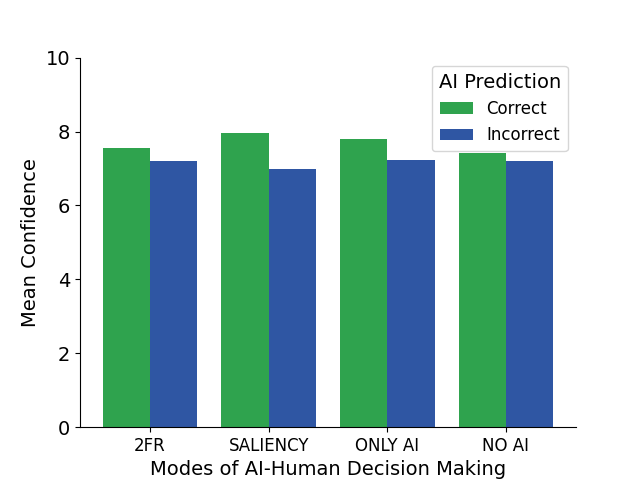} 
    \caption{Clinician confidence across modalities and AI correctness.}
    \label{fig:conf_modalities}
\end{figure}

\begin{figure}[t]
    \begin{subfigure}{
        \includegraphics[width=0.9\linewidth]{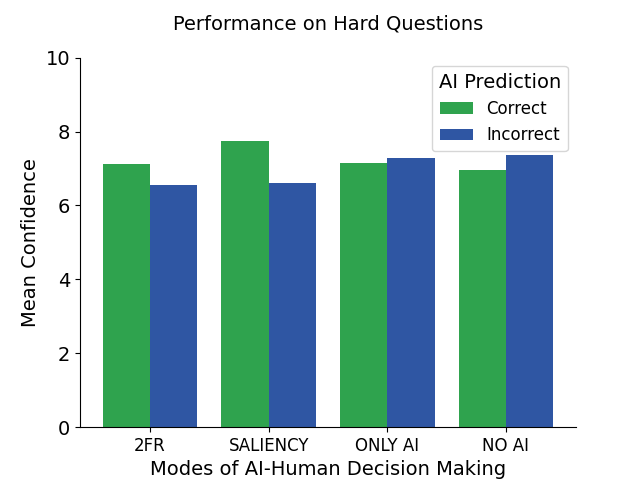}
    }
    \end{subfigure}
    \vfill   
    \begin{subfigure}{
        \includegraphics[width=0.9\linewidth]{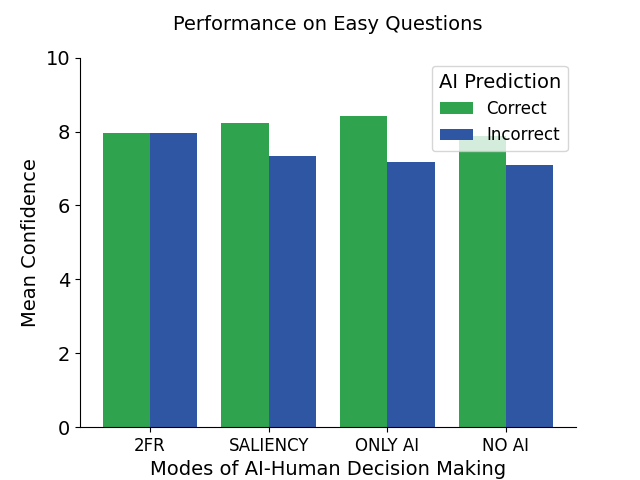}
    }
    \end{subfigure}
    \caption{Confidence across modalities and AI correctness based on chest x-ray difficulty.}
    \label{fig:conf_diff_modalities}
\end{figure}

A key observation in Fig. \ref{fig:conf_modalities} and \ref{fig:conf_diff_modalities} is that changes in clinician confidence are marginal, irrespective of AI correctness or the question's difficulty. For overall performance, clinician confidence remains relatively stable across modalities, with only slight differences between correct and incorrect AI predictions. This suggests that while AI correctness influences confidence to a small degree, its overall impact on clinician self-assurance is limited. When analyzing hard questions, clinician confidence is slightly lower compared to easy questions, particularly when AI predictions are incorrect. However, the differences are minimal, with 2FR and Saliency showing only small reductions in confidence. For easy questions, confidence remains uniformly high across all modalities, regardless of whether the AI prediction is correct, further emphasizing the marginal effect of AI correctness on confidence in simpler tasks. These findings highlight that clinician confidence is largely resilient to variations in AI correctness and task difficulty, with only slight shifts observed across conditions and modalities.

\subsection{Clinician Confidence and Accuracy}

\begin{figure}[t]
    \includegraphics[width=0.50\textwidth]{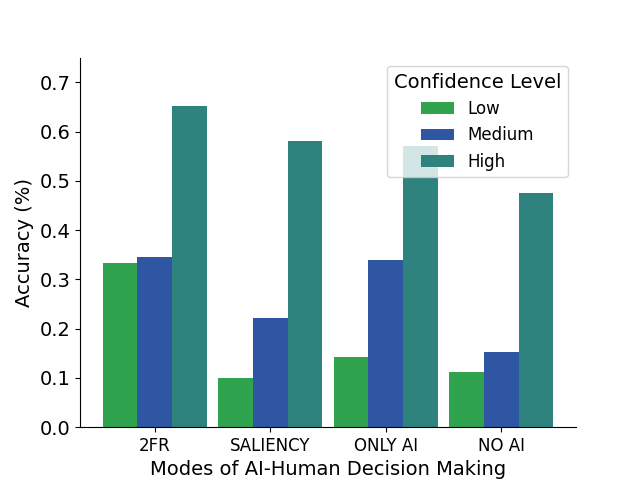}
    \caption{Clinician Accuracy Based On Confidence Levels.}
    \label{fig:acc_across_conf}
\end{figure}

Fig. \ref{fig:acc_across_conf} illustrates the accuracy across clinician confidence levels. Focusing on when clinicians exhibit low confidence (light green bars), 2FR achieved a moderate accuracy (~30\%). While this is lower than medium and high confidence groups, it remains the highest among all low-confidence results across modalities. 2FR achieves 3X more performance in low confidence when compared to Saliency(~10\%) and 2X when compared to Only AI (~15\%). This highlights the unique advantage of 2FR for individuals operating on questions they feel low confidence on. The 2FR approach likely aids participants by reinforcing diagnostic memory or offering explicit support, minimizing the cognitive burden experienced in uncertain scenarios. This could explain its significantly higher performance.

\section{Discussion}

It is tempting to believe that integrating AI-generated predictions into clinical workflows will inherently enhance diagnostic accuracy and efficiency. However, as with any technological advancement in critical domains like healthcare, it is imperative to empirically evaluate its actual impact on human decision-making. This is an important topic for the broader impact of AI and has been explored in the medical literature~\cite{gaube2021ai, jussupow2020we} but requires further study as human-AI systems are increasingly being integrated in healthcare. In this study, we conducted a rigorous investigation to assess how different modes of AI assistance influence clinicians' diagnostic performance and confidence when interpreting chest X-rays. We recruited 69 physicians across various specialties and levels of experience, and analyzing their responses to 12 diagnostic cases under different AI assistance modalities, we uncovered nuanced insights in AI-Human collaboration.

Our findings reveal that when AI predictions are correct, providing clinicians with additional explanatory features—such as 2FR examples of similar cases or saliency maps highlighting pertinent image regions—can enhance diagnostic accuracy compared to providing AI predictions alone or offering no AI assistance. Specifically, the 2FR modality, which presented AI diagnoses alongside representative images recognized by other physicians, resulted in the highest overall accuracy (70\%). This suggests that contextualizing AI outputs with relatable examples aids clinicians in better understanding and trusting AI recommendations. Conversely, when AI predictions were incorrect, clinician accuracy dropped significantly across all modalities to comparable levels if no AI predictions were given at all. This implies that when clinicians encounter questions that an AI fails on, they fall back to relying on their own expertise.

Interestingly, clinician confidence remained relatively stable across different AI modalities and was not significantly influenced by AI correctness or the difficulty level of the cases. This resilience in self-assessed confidence, despite fluctuations in actual diagnostic accuracy, points to a complex relationship between confidence and performance in AI-assisted decision-making. It suggests that clinicians may not adequately adjust their confidence levels in response to AI errors, which could lead to reduced vigilance in critical evaluation scenarios.

When clinicians exhibit low confidence, we observe that the 2FR modality improves AI-Human decision making accuracy. This has important implications for the design and implementation of AI decision support systems in medicine. Incorporating explanatory features that enhance interpretability can improve clinician performance when AI is accurate and when clinicians lack confidence in their response. A verification strategy like 2FR can increase overall performance of AI-Human systems, especially when a clinician does not feel confident in their decision.

\section{Conclusion}

This study shows how simple changes to AI decision making support systems that include a verification-based component can lead to improvements in clinician performance. The utility of our proposed method, `2FR', is not well explored in this domain, and we hope that our study will inspire a new line of research into improving this method from intelligently picking similar types of images to incorporating model uncertainty in how references images are presented.  While our study focused on chest X-ray interpretation, it would be valuable to extend this research to other diagnostic domains and complex clinical tasks. Investigating the long-term effects of AI assistance on clinician learning, diagnostic strategies, and patient outcomes will provide deeper insights into optimizing human-AI collaboration in healthcare. Addressing these areas is crucial to harnessing the full potential of AI while safeguarding the quality and integrity of medical decision-making. 

\bibliographystyle{ACM-Reference-Format}
\bibliography{sample-base}


\begin{thebibliography}{54}


\ifx \showCODEN    \undefined \def \showCODEN     #1{\unskip}     \fi
\ifx \showDOI      \undefined \def \showDOI       #1{#1}\fi
\ifx \showISBNx    \undefined \def \showISBNx     #1{\unskip}     \fi
\ifx \showISBNxiii \undefined \def \showISBNxiii  #1{\unskip}     \fi
\ifx \showISSN     \undefined \def \showISSN      #1{\unskip}     \fi
\ifx \showLCCN     \undefined \def \showLCCN      #1{\unskip}     \fi
\ifx \shownote     \undefined \def \shownote      #1{#1}          \fi
\ifx \showarticletitle \undefined \def \showarticletitle #1{#1}   \fi
\ifx \showURL      \undefined \def \showURL       {\relax}        \fi
\providecommand\bibfield[2]{#2}
\providecommand\bibinfo[2]{#2}
\providecommand\natexlab[1]{#1}
\providecommand\showeprint[2][]{arXiv:#2}

\bibitem[Alam and Mueller(2021)]%
        {alam2021examining}
\bibfield{author}{\bibinfo{person}{Lamia Alam} {and} \bibinfo{person}{Shane Mueller}.} \bibinfo{year}{2021}\natexlab{}.
\newblock \showarticletitle{Examining the effect of explanation on satisfaction and trust in AI diagnostic systems}.
\newblock \bibinfo{journal}{\emph{BMC medical informatics and decision making}} \bibinfo{volume}{21}, \bibinfo{number}{1} (\bibinfo{year}{2021}), \bibinfo{pages}{178}.
\newblock


\bibitem[Barata et~al\mbox{.}(2023)]%
        {barata2023reinforcement}
\bibfield{author}{\bibinfo{person}{Catarina Barata}, \bibinfo{person}{Veronica Rotemberg}, \bibinfo{person}{Noel~CF Codella}, \bibinfo{person}{Philipp Tschandl}, \bibinfo{person}{Christoph Rinner}, \bibinfo{person}{Bengu~Nisa Akay}, \bibinfo{person}{Zoe Apalla}, \bibinfo{person}{Giuseppe Argenziano}, \bibinfo{person}{Allan Halpern}, \bibinfo{person}{Aimilios Lallas}, {et~al\mbox{.}}} \bibinfo{year}{2023}\natexlab{}.
\newblock \showarticletitle{A reinforcement learning model for AI-based decision support in skin cancer}.
\newblock \bibinfo{journal}{\emph{Nature Medicine}} \bibinfo{volume}{29}, \bibinfo{number}{8} (\bibinfo{year}{2023}), \bibinfo{pages}{1941--1946}.
\newblock


\bibitem[Bilgic and Mooney(2005)]%
        {bilgic2005explaining}
\bibfield{author}{\bibinfo{person}{Mustafa Bilgic} {and} \bibinfo{person}{Raymond~J Mooney}.} \bibinfo{year}{2005}\natexlab{}.
\newblock \showarticletitle{Explaining recommendations: Satisfaction vs. promotion}. In \bibinfo{booktitle}{\emph{Beyond personalization workshop, IUI}}, Vol.~\bibinfo{volume}{5}. \bibinfo{pages}{153}.
\newblock


\bibitem[Bu{\c{c}}inca et~al\mbox{.}(2020)]%
        {buccinca2020proxy}
\bibfield{author}{\bibinfo{person}{Zana Bu{\c{c}}inca}, \bibinfo{person}{Phoebe Lin}, \bibinfo{person}{Krzysztof~Z Gajos}, {and} \bibinfo{person}{Elena~L Glassman}.} \bibinfo{year}{2020}\natexlab{}.
\newblock \showarticletitle{Proxy tasks and subjective measures can be misleading in evaluating explainable AI systems}. In \bibinfo{booktitle}{\emph{Proceedings of the 25th international conference on intelligent user interfaces}}. \bibinfo{pages}{454--464}.
\newblock


\bibitem[Bu{\c{c}}inca et~al\mbox{.}(2021)]%
        {buccinca2021trust}
\bibfield{author}{\bibinfo{person}{Zana Bu{\c{c}}inca}, \bibinfo{person}{Maja~Barbara Malaya}, {and} \bibinfo{person}{Krzysztof~Z Gajos}.} \bibinfo{year}{2021}\natexlab{}.
\newblock \showarticletitle{To trust or to think: cognitive forcing functions can reduce overreliance on AI in AI-assisted decision-making}.
\newblock \bibinfo{journal}{\emph{Proceedings of the ACM on Human-Computer Interaction}} \bibinfo{volume}{5}, \bibinfo{number}{CSCW1} (\bibinfo{year}{2021}), \bibinfo{pages}{1--21}.
\newblock


\bibitem[Bussone et~al\mbox{.}(2015)]%
        {bussone2015role}
\bibfield{author}{\bibinfo{person}{Adrian Bussone}, \bibinfo{person}{Simone Stumpf}, {and} \bibinfo{person}{Dympna O'Sullivan}.} \bibinfo{year}{2015}\natexlab{}.
\newblock \showarticletitle{The role of explanations on trust and reliance in clinical decision support systems}. In \bibinfo{booktitle}{\emph{2015 international conference on healthcare informatics}}. IEEE, \bibinfo{pages}{160--169}.
\newblock


\bibitem[Cai et~al\mbox{.}(2019)]%
        {cai2019hello}
\bibfield{author}{\bibinfo{person}{Carrie~J Cai}, \bibinfo{person}{Samantha Winter}, \bibinfo{person}{David Steiner}, \bibinfo{person}{Lauren Wilcox}, {and} \bibinfo{person}{Michael Terry}.} \bibinfo{year}{2019}\natexlab{}.
\newblock \showarticletitle{" Hello AI": uncovering the onboarding needs of medical practitioners for human-AI collaborative decision-making}.
\newblock \bibinfo{journal}{\emph{Proceedings of the ACM on Human-computer Interaction}} \bibinfo{volume}{3}, \bibinfo{number}{CSCW} (\bibinfo{year}{2019}), \bibinfo{pages}{1--24}.
\newblock


\bibitem[Devaraj et~al\mbox{.}(2014)]%
        {devaraj2014barriers}
\bibfield{author}{\bibinfo{person}{Srikant Devaraj}, \bibinfo{person}{Sushil~K Sharma}, \bibinfo{person}{Dyan~J Fausto}, \bibinfo{person}{Sara Viernes}, \bibinfo{person}{Hadi Kharrazi}, {et~al\mbox{.}}} \bibinfo{year}{2014}\natexlab{}.
\newblock \showarticletitle{Barriers and facilitators to clinical decision support systems adoption: a systematic review}.
\newblock \bibinfo{journal}{\emph{Journal of Business Administration Research}} \bibinfo{volume}{3}, \bibinfo{number}{2} (\bibinfo{year}{2014}), \bibinfo{pages}{36}.
\newblock


\bibitem[Ehsan et~al\mbox{.}(2021)]%
        {ehsan2021operationalizing}
\bibfield{author}{\bibinfo{person}{Upol Ehsan}, \bibinfo{person}{Philipp Wintersberger}, \bibinfo{person}{Q~Vera Liao}, \bibinfo{person}{Martina Mara}, \bibinfo{person}{Marc Streit}, \bibinfo{person}{Sandra Wachter}, \bibinfo{person}{Andreas Riener}, {and} \bibinfo{person}{Mark~O Riedl}.} \bibinfo{year}{2021}\natexlab{}.
\newblock \showarticletitle{Operationalizing human-centered perspectives in explainable AI}. In \bibinfo{booktitle}{\emph{Extended abstracts of the 2021 CHI conference on human factors in computing systems}}. \bibinfo{pages}{1--6}.
\newblock


\bibitem[Elwyn et~al\mbox{.}(2013)]%
        {elwyn2013many}
\bibfield{author}{\bibinfo{person}{Glyn Elwyn}, \bibinfo{person}{Isabelle Scholl}, \bibinfo{person}{Caroline Tietbohl}, \bibinfo{person}{Mala Mann}, \bibinfo{person}{Adrian~GK Edwards}, \bibinfo{person}{Catharine Clay}, \bibinfo{person}{France L{\'e}gar{\'e}}, \bibinfo{person}{Trudy van~der Weijden}, \bibinfo{person}{Carmen~L Lewis}, \bibinfo{person}{Richard~M Wexler}, {et~al\mbox{.}}} \bibinfo{year}{2013}\natexlab{}.
\newblock \showarticletitle{“Many miles to go…”: a systematic review of the implementation of patient decision support interventions into routine clinical practice}.
\newblock \bibinfo{journal}{\emph{BMC medical informatics and decision making}}  \bibinfo{volume}{13} (\bibinfo{year}{2013}), \bibinfo{pages}{1--10}.
\newblock


\bibitem[Feldman et~al\mbox{.}(2019)]%
        {feldman2019artificial}
\bibfield{author}{\bibinfo{person}{Robin~C Feldman}, \bibinfo{person}{Ehrik Aldana}, {and} \bibinfo{person}{Kara Stein}.} \bibinfo{year}{2019}\natexlab{}.
\newblock \showarticletitle{Artificial intelligence in the health care space: how we can trust what we cannot know}.
\newblock \bibinfo{journal}{\emph{Stan. L. \& Pol'y Rev.}}  \bibinfo{volume}{30} (\bibinfo{year}{2019}), \bibinfo{pages}{399}.
\newblock


\bibitem[Fok and Weld(2023)]%
        {fok2023search}
\bibfield{author}{\bibinfo{person}{Raymond Fok} {and} \bibinfo{person}{Daniel~S Weld}.} \bibinfo{year}{2023}\natexlab{}.
\newblock \showarticletitle{In search of verifiability: Explanations rarely enable complementary performance in ai-advised decision making}.
\newblock \bibinfo{journal}{\emph{arXiv preprint arXiv:2305.07722}} (\bibinfo{year}{2023}).
\newblock


\bibitem[Gaube et~al\mbox{.}(2023)]%
        {gaube2023non}
\bibfield{author}{\bibinfo{person}{Susanne Gaube}, \bibinfo{person}{Harini Suresh}, \bibinfo{person}{Martina Raue}, \bibinfo{person}{Eva Lermer}, \bibinfo{person}{Timo~K Koch}, \bibinfo{person}{Matthias~FC Hudecek}, \bibinfo{person}{Alun~D Ackery}, \bibinfo{person}{Samir~C Grover}, \bibinfo{person}{Joseph~F Coughlin}, \bibinfo{person}{Dieter Frey}, {et~al\mbox{.}}} \bibinfo{year}{2023}\natexlab{}.
\newblock \showarticletitle{Non-task expert physicians benefit from correct explainable AI advice when reviewing X-rays}.
\newblock \bibinfo{journal}{\emph{Scientific reports}} \bibinfo{volume}{13}, \bibinfo{number}{1} (\bibinfo{year}{2023}), \bibinfo{pages}{1383}.
\newblock


\bibitem[Gaube et~al\mbox{.}(2021)]%
        {gaube2021ai}
\bibfield{author}{\bibinfo{person}{Susanne Gaube}, \bibinfo{person}{Harini Suresh}, \bibinfo{person}{Martina Raue}, \bibinfo{person}{Alexander Merritt}, \bibinfo{person}{Seth~J Berkowitz}, \bibinfo{person}{Eva Lermer}, \bibinfo{person}{Joseph~F Coughlin}, \bibinfo{person}{John~V Guttag}, \bibinfo{person}{Errol Colak}, {and} \bibinfo{person}{Marzyeh Ghassemi}.} \bibinfo{year}{2021}\natexlab{}.
\newblock \showarticletitle{Do as AI say: susceptibility in deployment of clinical decision-aids}.
\newblock \bibinfo{journal}{\emph{NPJ digital medicine}} \bibinfo{volume}{4}, \bibinfo{number}{1} (\bibinfo{year}{2021}), \bibinfo{pages}{31}.
\newblock


\bibitem[Ghai et~al\mbox{.}(2021)]%
        {ghai2021explainable}
\bibfield{author}{\bibinfo{person}{Bhavya Ghai}, \bibinfo{person}{Q~Vera Liao}, \bibinfo{person}{Yunfeng Zhang}, \bibinfo{person}{Rachel Bellamy}, {and} \bibinfo{person}{Klaus Mueller}.} \bibinfo{year}{2021}\natexlab{}.
\newblock \showarticletitle{Explainable active learning (xal) toward ai explanations as interfaces for machine teachers}.
\newblock \bibinfo{journal}{\emph{Proceedings of the ACM on Human-Computer Interaction}} \bibinfo{volume}{4}, \bibinfo{number}{CSCW3} (\bibinfo{year}{2021}), \bibinfo{pages}{1--28}.
\newblock


\bibitem[Ghassemi et~al\mbox{.}(2021)]%
        {ghassemi2021false}
\bibfield{author}{\bibinfo{person}{Marzyeh Ghassemi}, \bibinfo{person}{Luke Oakden-Rayner}, {and} \bibinfo{person}{Andrew~L Beam}.} \bibinfo{year}{2021}\natexlab{}.
\newblock \showarticletitle{The false hope of current approaches to explainable artificial intelligence in health care}.
\newblock \bibinfo{journal}{\emph{The Lancet Digital Health}} \bibinfo{volume}{3}, \bibinfo{number}{11} (\bibinfo{year}{2021}), \bibinfo{pages}{e745--e750}.
\newblock


\bibitem[Gombolay et~al\mbox{.}(2024)]%
        {gombolay2024effects}
\bibfield{author}{\bibinfo{person}{Grace~Y Gombolay}, \bibinfo{person}{Andrew Silva}, \bibinfo{person}{Mariah Schrum}, \bibinfo{person}{Nakul Gopalan}, \bibinfo{person}{Jamika Hallman-Cooper}, \bibinfo{person}{Monideep Dutt}, {and} \bibinfo{person}{Matthew Gombolay}.} \bibinfo{year}{2024}\natexlab{}.
\newblock \showarticletitle{Effects of explainable artificial intelligence in neurology decision support}.
\newblock \bibinfo{journal}{\emph{Annals of Clinical and Translational Neurology}} \bibinfo{volume}{11}, \bibinfo{number}{5} (\bibinfo{year}{2024}), \bibinfo{pages}{1224--1235}.
\newblock


\bibitem[Groh et~al\mbox{.}(2024)]%
        {groh2024deep}
\bibfield{author}{\bibinfo{person}{Matthew Groh}, \bibinfo{person}{Omar Badri}, \bibinfo{person}{Roxana Daneshjou}, \bibinfo{person}{Arash Koochek}, \bibinfo{person}{Caleb Harris}, \bibinfo{person}{Luis~R Soenksen}, \bibinfo{person}{P~Murali Doraiswamy}, {and} \bibinfo{person}{Rosalind Picard}.} \bibinfo{year}{2024}\natexlab{}.
\newblock \showarticletitle{Deep learning-aided decision support for diagnosis of skin disease across skin tones}.
\newblock \bibinfo{journal}{\emph{Nature Medicine}} \bibinfo{volume}{30}, \bibinfo{number}{2} (\bibinfo{year}{2024}), \bibinfo{pages}{573--583}.
\newblock


\bibitem[Guo et~al\mbox{.}(2024)]%
        {guo2024improve}
\bibfield{author}{\bibinfo{person}{Yingxuan Guo}, \bibinfo{person}{Changke Huang}, \bibinfo{person}{Yaying Sheng}, \bibinfo{person}{Wenjie Zhang}, \bibinfo{person}{Xin Ye}, \bibinfo{person}{Hengli Lian}, \bibinfo{person}{Jiahao Xu}, {and} \bibinfo{person}{Yiqi Chen}.} \bibinfo{year}{2024}\natexlab{}.
\newblock \showarticletitle{Improve the efficiency and accuracy of ophthalmologists’ clinical decision-making based on AI technology}.
\newblock \bibinfo{journal}{\emph{BMC Medical Informatics and Decision Making}} \bibinfo{volume}{24}, \bibinfo{number}{1} (\bibinfo{year}{2024}), \bibinfo{pages}{192}.
\newblock


\bibitem[Hekler et~al\mbox{.}(2019)]%
        {hekler2019superior}
\bibfield{author}{\bibinfo{person}{Achim Hekler}, \bibinfo{person}{Jochen~S Utikal}, \bibinfo{person}{Alexander~H Enk}, \bibinfo{person}{Axel Hauschild}, \bibinfo{person}{Michael Weichenthal}, \bibinfo{person}{Roman~C Maron}, \bibinfo{person}{Carola Berking}, \bibinfo{person}{Sebastian Haferkamp}, \bibinfo{person}{Joachim Klode}, \bibinfo{person}{Dirk Schadendorf}, {et~al\mbox{.}}} \bibinfo{year}{2019}\natexlab{}.
\newblock \showarticletitle{Superior skin cancer classification by the combination of human and artificial intelligence}.
\newblock \bibinfo{journal}{\emph{European Journal of Cancer}}  \bibinfo{volume}{120} (\bibinfo{year}{2019}), \bibinfo{pages}{114--121}.
\newblock


\bibitem[Henry et~al\mbox{.}(2022)]%
        {henry2022human}
\bibfield{author}{\bibinfo{person}{Katharine~E Henry}, \bibinfo{person}{Rachel Kornfield}, \bibinfo{person}{Anirudh Sridharan}, \bibinfo{person}{Robert~C Linton}, \bibinfo{person}{Catherine Groh}, \bibinfo{person}{Tony Wang}, \bibinfo{person}{Albert Wu}, \bibinfo{person}{Bilge Mutlu}, {and} \bibinfo{person}{Suchi Saria}.} \bibinfo{year}{2022}\natexlab{}.
\newblock \showarticletitle{Human--machine teaming is key to AI adoption: clinicians’ experiences with a deployed machine learning system}.
\newblock \bibinfo{journal}{\emph{NPJ digital medicine}} \bibinfo{volume}{5}, \bibinfo{number}{1} (\bibinfo{year}{2022}), \bibinfo{pages}{97}.
\newblock


\bibitem[Horne et~al\mbox{.}(2019)]%
        {horne2019rating}
\bibfield{author}{\bibinfo{person}{Benjamin~D Horne}, \bibinfo{person}{Dorit Nevo}, \bibinfo{person}{John O’Donovan}, \bibinfo{person}{Jin-Hee Cho}, {and} \bibinfo{person}{Sibel Adal{\i}}.} \bibinfo{year}{2019}\natexlab{}.
\newblock \showarticletitle{Rating reliability and bias in news articles: Does AI assistance help everyone?}. In \bibinfo{booktitle}{\emph{Proceedings of the International AAAI Conference on Web and Social Media}}, Vol.~\bibinfo{volume}{13}. \bibinfo{pages}{247--256}.
\newblock


\bibitem[Jacobs et~al\mbox{.}(2021)]%
        {jacobs2021machine}
\bibfield{author}{\bibinfo{person}{Maia Jacobs}, \bibinfo{person}{Melanie~F Pradier}, \bibinfo{person}{Thomas~H McCoy~Jr}, \bibinfo{person}{Roy~H Perlis}, \bibinfo{person}{Finale Doshi-Velez}, {and} \bibinfo{person}{Krzysztof~Z Gajos}.} \bibinfo{year}{2021}\natexlab{}.
\newblock \showarticletitle{How machine-learning recommendations influence clinician treatment selections: the example of antidepressant selection}.
\newblock \bibinfo{journal}{\emph{Translational psychiatry}} \bibinfo{volume}{11}, \bibinfo{number}{1} (\bibinfo{year}{2021}), \bibinfo{pages}{108}.
\newblock


\bibitem[Jussupow et~al\mbox{.}(2020)]%
        {jussupow2020we}
\bibfield{author}{\bibinfo{person}{Ekaterina Jussupow}, \bibinfo{person}{Izak Benbasat}, {and} \bibinfo{person}{Armin Heinzl}.} \bibinfo{year}{2020}\natexlab{}.
\newblock \showarticletitle{Why are we averse towards algorithms? A comprehensive literature review on algorithm aversion}.
\newblock  (\bibinfo{year}{2020}).
\newblock


\bibitem[Khosravi et~al\mbox{.}(2024)]%
        {khosravi2024artificial}
\bibfield{author}{\bibinfo{person}{Mohsen Khosravi}, \bibinfo{person}{Zahra Zare}, \bibinfo{person}{Seyyed~Morteza Mojtabaeian}, {and} \bibinfo{person}{Reyhane Izadi}.} \bibinfo{year}{2024}\natexlab{}.
\newblock \showarticletitle{Artificial intelligence and decision-making in healthcare: a thematic analysis of a systematic review of reviews}.
\newblock \bibinfo{journal}{\emph{Health services research and managerial epidemiology}}  \bibinfo{volume}{11} (\bibinfo{year}{2024}), \bibinfo{pages}{23333928241234863}.
\newblock


\bibitem[Kim et~al\mbox{.}(2022)]%
        {kim2022hive}
\bibfield{author}{\bibinfo{person}{Sunnie~SY Kim}, \bibinfo{person}{Nicole Meister}, \bibinfo{person}{Vikram~V Ramaswamy}, \bibinfo{person}{Ruth Fong}, {and} \bibinfo{person}{Olga Russakovsky}.} \bibinfo{year}{2022}\natexlab{}.
\newblock \showarticletitle{HIVE: Evaluating the human interpretability of visual explanations}. In \bibinfo{booktitle}{\emph{European Conference on Computer Vision}}. Springer, \bibinfo{pages}{280--298}.
\newblock


\bibitem[Lai et~al\mbox{.}(2020)]%
        {lai2020chicago}
\bibfield{author}{\bibinfo{person}{Vivian Lai}, \bibinfo{person}{Han Liu}, {and} \bibinfo{person}{Chenhao Tan}.} \bibinfo{year}{2020}\natexlab{}.
\newblock \showarticletitle{" Why is' Chicago'deceptive?" Towards Building Model-Driven Tutorials for Humans}. In \bibinfo{booktitle}{\emph{Proceedings of the 2020 CHI Conference on Human Factors in Computing Systems}}. \bibinfo{pages}{1--13}.
\newblock


\bibitem[Lai and Tan(2019)]%
        {lai2019human}
\bibfield{author}{\bibinfo{person}{Vivian Lai} {and} \bibinfo{person}{Chenhao Tan}.} \bibinfo{year}{2019}\natexlab{}.
\newblock \showarticletitle{On human predictions with explanations and predictions of machine learning models: A case study on deception detection}. In \bibinfo{booktitle}{\emph{Proceedings of the conference on fairness, accountability, and transparency}}. \bibinfo{pages}{29--38}.
\newblock


\bibitem[Li et~al\mbox{.}(2023)]%
        {li2023artificial}
\bibfield{author}{\bibinfo{person}{Zhongwen Li}, \bibinfo{person}{Lei Wang}, \bibinfo{person}{Xuefang Wu}, \bibinfo{person}{Jiewei Jiang}, \bibinfo{person}{Wei Qiang}, \bibinfo{person}{He Xie}, \bibinfo{person}{Hongjian Zhou}, \bibinfo{person}{Shanjun Wu}, \bibinfo{person}{Yi Shao}, {and} \bibinfo{person}{Wei Chen}.} \bibinfo{year}{2023}\natexlab{}.
\newblock \showarticletitle{Artificial intelligence in ophthalmology: The path to the real-world clinic}.
\newblock \bibinfo{journal}{\emph{Cell Reports Medicine}} \bibinfo{volume}{4}, \bibinfo{number}{7} (\bibinfo{year}{2023}).
\newblock


\bibitem[Massalha et~al\mbox{.}(2018)]%
        {massalha2018decision}
\bibfield{author}{\bibinfo{person}{Samia Massalha}, \bibinfo{person}{Owen Clarkin}, \bibinfo{person}{Rebecca Thornhill}, \bibinfo{person}{Glenn Wells}, {and} \bibinfo{person}{Benjamin~JW Chow}.} \bibinfo{year}{2018}\natexlab{}.
\newblock \showarticletitle{Decision support tools, systems, and artificial intelligence in cardiac imaging}.
\newblock \bibinfo{journal}{\emph{Canadian Journal of Cardiology}} \bibinfo{volume}{34}, \bibinfo{number}{7} (\bibinfo{year}{2018}), \bibinfo{pages}{827--838}.
\newblock


\bibitem[Miller(2019)]%
        {miller2019explanation}
\bibfield{author}{\bibinfo{person}{Tim Miller}.} \bibinfo{year}{2019}\natexlab{}.
\newblock \showarticletitle{Explanation in artificial intelligence: Insights from the social sciences}.
\newblock \bibinfo{journal}{\emph{Artificial intelligence}}  \bibinfo{volume}{267} (\bibinfo{year}{2019}), \bibinfo{pages}{1--38}.
\newblock


\bibitem[Naiseh et~al\mbox{.}(2023)]%
        {naiseh2023different}
\bibfield{author}{\bibinfo{person}{Mohammad Naiseh}, \bibinfo{person}{Dena Al-Thani}, \bibinfo{person}{Nan Jiang}, {and} \bibinfo{person}{Raian Ali}.} \bibinfo{year}{2023}\natexlab{}.
\newblock \showarticletitle{How the different explanation classes impact trust calibration: The case of clinical decision support systems}.
\newblock \bibinfo{journal}{\emph{International Journal of Human-Computer Studies}}  \bibinfo{volume}{169} (\bibinfo{year}{2023}), \bibinfo{pages}{102941}.
\newblock


\bibitem[Nourani et~al\mbox{.}(2021)]%
        {nourani2021anchoring}
\bibfield{author}{\bibinfo{person}{Mahsan Nourani}, \bibinfo{person}{Chiradeep Roy}, \bibinfo{person}{Jeremy~E Block}, \bibinfo{person}{Donald~R Honeycutt}, \bibinfo{person}{Tahrima Rahman}, \bibinfo{person}{Eric Ragan}, {and} \bibinfo{person}{Vibhav Gogate}.} \bibinfo{year}{2021}\natexlab{}.
\newblock \showarticletitle{Anchoring bias affects mental model formation and user reliance in explainable ai systems}. In \bibinfo{booktitle}{\emph{26th International Conference on Intelligent User Interfaces}}. \bibinfo{pages}{340--350}.
\newblock


\bibitem[Olawade et~al\mbox{.}(2024)]%
        {olawade2024advancements}
\bibfield{author}{\bibinfo{person}{David~B Olawade}, \bibinfo{person}{Nicholas Aderinto}, \bibinfo{person}{Gbolahan Olatunji}, \bibinfo{person}{Emmanuel Kokori}, \bibinfo{person}{Aanuoluwapo~C David-Olawade}, {and} \bibinfo{person}{Manizha Hadi}.} \bibinfo{year}{2024}\natexlab{}.
\newblock \showarticletitle{Advancements and applications of Artificial Intelligence in cardiology: Current trends and future prospects}.
\newblock \bibinfo{journal}{\emph{Journal of Medicine, Surgery, and Public Health}} (\bibinfo{year}{2024}), \bibinfo{pages}{100109}.
\newblock


\bibitem[Rajpurkar(2017)]%
        {rajpurkar2017chexnet}
\bibfield{author}{\bibinfo{person}{P Rajpurkar}.} \bibinfo{year}{2017}\natexlab{}.
\newblock \showarticletitle{CheXNet: Radiologist-Level Pneumonia Detection on Chest X-Rays with Deep Learning}.
\newblock \bibinfo{journal}{\emph{ArXiv abs/1711}}  \bibinfo{volume}{5225} (\bibinfo{year}{2017}).
\newblock


\bibitem[Ribeiro et~al\mbox{.}(2016)]%
        {ribeiro2016should}
\bibfield{author}{\bibinfo{person}{Marco~Tulio Ribeiro}, \bibinfo{person}{Sameer Singh}, {and} \bibinfo{person}{Carlos Guestrin}.} \bibinfo{year}{2016}\natexlab{}.
\newblock \showarticletitle{" Why should i trust you?" Explaining the predictions of any classifier}. In \bibinfo{booktitle}{\emph{Proceedings of the 22nd ACM SIGKDD international conference on knowledge discovery and data mining}}. \bibinfo{pages}{1135--1144}.
\newblock


\bibitem[Schaekermann et~al\mbox{.}(2020)]%
        {schaekermann2020ambiguity}
\bibfield{author}{\bibinfo{person}{Mike Schaekermann}, \bibinfo{person}{Graeme Beaton}, \bibinfo{person}{Elaheh Sanoubari}, \bibinfo{person}{Andrew Lim}, \bibinfo{person}{Kate Larson}, {and} \bibinfo{person}{Edith Law}.} \bibinfo{year}{2020}\natexlab{}.
\newblock \showarticletitle{Ambiguity-aware ai assistants for medical data analysis}. In \bibinfo{booktitle}{\emph{Proceedings of the 2020 CHI conference on human factors in computing systems}}. \bibinfo{pages}{1--14}.
\newblock


\bibitem[Scheetz et~al\mbox{.}(2021)]%
        {scheetz2021survey}
\bibfield{author}{\bibinfo{person}{Jane Scheetz}, \bibinfo{person}{Philip Rothschild}, \bibinfo{person}{Myra McGuinness}, \bibinfo{person}{Xavier Hadoux}, \bibinfo{person}{H~Peter Soyer}, \bibinfo{person}{Monika Janda}, \bibinfo{person}{James~JJ Condon}, \bibinfo{person}{Luke Oakden-Rayner}, \bibinfo{person}{Lyle~J Palmer}, \bibinfo{person}{Stuart Keel}, {et~al\mbox{.}}} \bibinfo{year}{2021}\natexlab{}.
\newblock \showarticletitle{A survey of clinicians on the use of artificial intelligence in ophthalmology, dermatology, radiology and radiation oncology}.
\newblock \bibinfo{journal}{\emph{Scientific reports}} \bibinfo{volume}{11}, \bibinfo{number}{1} (\bibinfo{year}{2021}), \bibinfo{pages}{5193}.
\newblock


\bibitem[Selvaraju et~al\mbox{.}(2017)]%
        {selvaraju2017grad}
\bibfield{author}{\bibinfo{person}{Ramprasaath~R Selvaraju}, \bibinfo{person}{Michael Cogswell}, \bibinfo{person}{Abhishek Das}, \bibinfo{person}{Ramakrishna Vedantam}, \bibinfo{person}{Devi Parikh}, {and} \bibinfo{person}{Dhruv Batra}.} \bibinfo{year}{2017}\natexlab{}.
\newblock \showarticletitle{Grad-cam: Visual explanations from deep networks via gradient-based localization}. In \bibinfo{booktitle}{\emph{Proceedings of the IEEE international conference on computer vision}}. \bibinfo{pages}{618--626}.
\newblock


\bibitem[Shahtalebi et~al\mbox{.}(2021)]%
        {shahtalebi2021deep}
\bibfield{author}{\bibinfo{person}{Soroosh Shahtalebi}, \bibinfo{person}{S~Farokh Atashzar}, \bibinfo{person}{Rajni~V Patel}, \bibinfo{person}{Mandar~S Jog}, {and} \bibinfo{person}{Arash Mohammadi}.} \bibinfo{year}{2021}\natexlab{}.
\newblock \showarticletitle{A deep explainable artificial intelligent framework for neurological disorders discrimination}.
\newblock \bibinfo{journal}{\emph{Scientific reports}} \bibinfo{volume}{11}, \bibinfo{number}{1} (\bibinfo{year}{2021}), \bibinfo{pages}{9630}.
\newblock


\bibitem[Si et~al\mbox{.}(2023)]%
        {si2023large}
\bibfield{author}{\bibinfo{person}{Chenglei Si}, \bibinfo{person}{Navita Goyal}, \bibinfo{person}{Sherry~Tongshuang Wu}, \bibinfo{person}{Chen Zhao}, \bibinfo{person}{Shi Feng}, \bibinfo{person}{Hal Daum{\'e}~III}, {and} \bibinfo{person}{Jordan Boyd-Graber}.} \bibinfo{year}{2023}\natexlab{}.
\newblock \showarticletitle{Large Language Models Help Humans Verify Truthfulness--Except When They Are Convincingly Wrong}.
\newblock \bibinfo{journal}{\emph{arXiv preprint arXiv:2310.12558}} (\bibinfo{year}{2023}).
\newblock


\bibitem[Sivaraman et~al\mbox{.}(2023)]%
        {sivaraman2023ignore}
\bibfield{author}{\bibinfo{person}{Venkatesh Sivaraman}, \bibinfo{person}{Leigh~A Bukowski}, \bibinfo{person}{Joel Levin}, \bibinfo{person}{Jeremy~M Kahn}, {and} \bibinfo{person}{Adam Perer}.} \bibinfo{year}{2023}\natexlab{}.
\newblock \showarticletitle{Ignore, trust, or negotiate: Understanding clinician acceptance of AI-based treatment recommendations in health care}. In \bibinfo{booktitle}{\emph{Proceedings of the 2023 CHI Conference on Human Factors in Computing Systems}}. \bibinfo{pages}{1--18}.
\newblock


\bibitem[Sultanum et~al\mbox{.}(2018)]%
        {sultanum2018more}
\bibfield{author}{\bibinfo{person}{Nicole Sultanum}, \bibinfo{person}{Michael Brudno}, \bibinfo{person}{Daniel Wigdor}, {and} \bibinfo{person}{Fanny Chevalier}.} \bibinfo{year}{2018}\natexlab{}.
\newblock \showarticletitle{More text please! understanding and supporting the use of visualization for clinical text overview}. In \bibinfo{booktitle}{\emph{Proceedings of the 2018 CHI conference on human factors in computing systems}}. \bibinfo{pages}{1--13}.
\newblock


\bibitem[Tait et~al\mbox{.}(2010)]%
        {tait2010effect}
\bibfield{author}{\bibinfo{person}{Alan~R Tait}, \bibinfo{person}{Terri Voepel-Lewis}, \bibinfo{person}{Brian~J Zikmund-Fisher}, {and} \bibinfo{person}{Angela Fagerlin}.} \bibinfo{year}{2010}\natexlab{}.
\newblock \showarticletitle{The effect of format on parents' understanding of the risks and benefits of clinical research: a comparison between text, tables, and graphics}.
\newblock \bibinfo{journal}{\emph{Journal of health communication}} \bibinfo{volume}{15}, \bibinfo{number}{5} (\bibinfo{year}{2010}), \bibinfo{pages}{487--501}.
\newblock


\bibitem[Timmermans et~al\mbox{.}(2004)]%
        {timmermans2004different}
\bibfield{author}{\bibinfo{person}{Danielle Timmermans}, \bibinfo{person}{Bert Molewijk}, \bibinfo{person}{Anne Stiggelbout}, {and} \bibinfo{person}{Job Kievit}.} \bibinfo{year}{2004}\natexlab{}.
\newblock \showarticletitle{Different formats for communicating surgical risks to patients and the effect on choice of treatment}.
\newblock \bibinfo{journal}{\emph{Patient education and counseling}} \bibinfo{volume}{54}, \bibinfo{number}{3} (\bibinfo{year}{2004}), \bibinfo{pages}{255--263}.
\newblock


\bibitem[Tschandl et~al\mbox{.}(2020)]%
        {tschandl2020human}
\bibfield{author}{\bibinfo{person}{Philipp Tschandl}, \bibinfo{person}{Christoph Rinner}, \bibinfo{person}{Zoe Apalla}, \bibinfo{person}{Giuseppe Argenziano}, \bibinfo{person}{Noel Codella}, \bibinfo{person}{Allan Halpern}, \bibinfo{person}{Monika Janda}, \bibinfo{person}{Aimilios Lallas}, \bibinfo{person}{Caterina Longo}, \bibinfo{person}{Josep Malvehy}, {et~al\mbox{.}}} \bibinfo{year}{2020}\natexlab{}.
\newblock \showarticletitle{Human--computer collaboration for skin cancer recognition}.
\newblock \bibinfo{journal}{\emph{Nature medicine}} \bibinfo{volume}{26}, \bibinfo{number}{8} (\bibinfo{year}{2020}), \bibinfo{pages}{1229--1234}.
\newblock


\bibitem[Vasconcelos et~al\mbox{.}(2023)]%
        {vasconcelos2023explanations}
\bibfield{author}{\bibinfo{person}{Helena Vasconcelos}, \bibinfo{person}{Matthew J{\"o}rke}, \bibinfo{person}{Madeleine Grunde-McLaughlin}, \bibinfo{person}{Tobias Gerstenberg}, \bibinfo{person}{Michael~S Bernstein}, {and} \bibinfo{person}{Ranjay Krishna}.} \bibinfo{year}{2023}\natexlab{}.
\newblock \showarticletitle{Explanations can reduce overreliance on ai systems during decision-making}.
\newblock \bibinfo{journal}{\emph{Proceedings of the ACM on Human-Computer Interaction}} \bibinfo{volume}{7}, \bibinfo{number}{CSCW1} (\bibinfo{year}{2023}), \bibinfo{pages}{1--38}.
\newblock


\bibitem[Wang et~al\mbox{.}(2019)]%
        {wang2019designing}
\bibfield{author}{\bibinfo{person}{Danding Wang}, \bibinfo{person}{Qian Yang}, \bibinfo{person}{Ashraf Abdul}, {and} \bibinfo{person}{Brian~Y Lim}.} \bibinfo{year}{2019}\natexlab{}.
\newblock \showarticletitle{Designing theory-driven user-centric explainable AI}. In \bibinfo{booktitle}{\emph{Proceedings of the 2019 CHI conference on human factors in computing systems}}. \bibinfo{pages}{1--15}.
\newblock


\bibitem[Wang et~al\mbox{.}(2017)]%
        {wang2017chestx}
\bibfield{author}{\bibinfo{person}{Xiaosong Wang}, \bibinfo{person}{Yifan Peng}, \bibinfo{person}{Le Lu}, \bibinfo{person}{Zhiyong Lu}, \bibinfo{person}{Mohammadhadi Bagheri}, {and} \bibinfo{person}{Ronald~M Summers}.} \bibinfo{year}{2017}\natexlab{}.
\newblock \showarticletitle{Chestx-ray8: Hospital-scale chest x-ray database and benchmarks on weakly-supervised classification and localization of common thorax diseases}. In \bibinfo{booktitle}{\emph{Proceedings of the IEEE conference on computer vision and pattern recognition}}. \bibinfo{pages}{2097--2106}.
\newblock


\bibitem[Wyatt and Altman(1995)]%
        {wyatt1995commentary}
\bibfield{author}{\bibinfo{person}{Jeremy~C Wyatt} {and} \bibinfo{person}{Douglas~G Altman}.} \bibinfo{year}{1995}\natexlab{}.
\newblock \showarticletitle{Commentary: Prognostic models: clinically useful or quickly forgotten?}
\newblock \bibinfo{journal}{\emph{Bmj}} \bibinfo{volume}{311}, \bibinfo{number}{7019} (\bibinfo{year}{1995}), \bibinfo{pages}{1539--1541}.
\newblock


\bibitem[Xie et~al\mbox{.}(2020)]%
        {xie2020chexplain}
\bibfield{author}{\bibinfo{person}{Yao Xie}, \bibinfo{person}{Melody Chen}, \bibinfo{person}{David Kao}, \bibinfo{person}{Ge Gao}, {and} \bibinfo{person}{Xiang'Anthony' Chen}.} \bibinfo{year}{2020}\natexlab{}.
\newblock \showarticletitle{CheXplain: enabling physicians to explore and understand data-driven, AI-enabled medical imaging analysis}. In \bibinfo{booktitle}{\emph{Proceedings of the 2020 CHI Conference on Human Factors in Computing Systems}}. \bibinfo{pages}{1--13}.
\newblock


\bibitem[Yang et~al\mbox{.}(2019)]%
        {yang2019unremarkable}
\bibfield{author}{\bibinfo{person}{Qian Yang}, \bibinfo{person}{Aaron Steinfeld}, {and} \bibinfo{person}{John Zimmerman}.} \bibinfo{year}{2019}\natexlab{}.
\newblock \showarticletitle{Unremarkable AI: Fitting intelligent decision support into critical, clinical decision-making processes}. In \bibinfo{booktitle}{\emph{Proceedings of the 2019 CHI conference on human factors in computing systems}}. \bibinfo{pages}{1--11}.
\newblock


\bibitem[Yu et~al\mbox{.}(2024)]%
        {yu2024heterogeneity}
\bibfield{author}{\bibinfo{person}{Feiyang Yu}, \bibinfo{person}{Alex Moehring}, \bibinfo{person}{Oishi Banerjee}, \bibinfo{person}{Tobias Salz}, \bibinfo{person}{Nikhil Agarwal}, {and} \bibinfo{person}{Pranav Rajpurkar}.} \bibinfo{year}{2024}\natexlab{}.
\newblock \showarticletitle{Heterogeneity and predictors of the effects of AI assistance on radiologists}.
\newblock \bibinfo{journal}{\emph{Nature Medicine}} \bibinfo{volume}{30}, \bibinfo{number}{3} (\bibinfo{year}{2024}), \bibinfo{pages}{837--849}.
\newblock


\bibitem[Zytek et~al\mbox{.}(2021)]%
        {zytek2021sibyl}
\bibfield{author}{\bibinfo{person}{Alexandra Zytek}, \bibinfo{person}{Dongyu Liu}, \bibinfo{person}{Rhema Vaithianathan}, {and} \bibinfo{person}{Kalyan Veeramachaneni}.} \bibinfo{year}{2021}\natexlab{}.
\newblock \showarticletitle{Sibyl: Understanding and addressing the usability challenges of machine learning in high-stakes decision making}.
\newblock \bibinfo{journal}{\emph{IEEE Transactions on Visualization and Computer Graphics}} \bibinfo{volume}{28}, \bibinfo{number}{1} (\bibinfo{year}{2021}), \bibinfo{pages}{1161--1171}.
\newblock


\end{thebibliography}

\end{document}